\documentclass[aps,prd,superscriptaddress,long,notitlepage,balancelastpage,nofootinbib,floatfix,twocolumn, preprintnumbers]{revtex4-1}
\pdfoutput=1
\usepackage{amsmath,mathtools,amssymb,amsthm,amsxtra,overpic,bbm,epsfig, subfigure,url, bm}
\usepackage{hyperref}
\usepackage{mathrsfs}
\usepackage{color,xcolor}
\usepackage{comment}
\usepackage{float}
\usepackage{enumitem}
\usepackage{slashed}
\usepackage{multirow}
\usepackage{appendix}
\usepackage[normalem]{ulem}
\usepackage{wasysym}

\DeclareUnicodeCharacter{03BC}{\ensuremath{\mu}}
\DeclareUnicodeCharacter{2032}{\ensuremath{'}}
\definecolor{nicered}{rgb}{0.5,0.,0.}
\definecolor{nicegreen}{rgb}{0.,0.5,0.}
\definecolor{niceblue}{rgb}{0.,0.,0.5}
\hypersetup{
	colorlinks=true,
	linkcolor=black,
	filecolor=nicegreen,      
	urlcolor=niceblue,
	citecolor=nicered,
}

\makeatletter
\newcommand*{\balancecolsandclearpage}{%
	\close@column@grid
	\cleardoublepage
	\twocolumngrid
}

\newcommand{\Refe}{Ref.}

\begin{document}

\preprint{HRI-RECAPP-2026-07}

\title{\vspace{1cm} \large 
Probing Neutrinophilic Long-range Forces  at DUNE 
}
\author{\bf Sudip Jana}
\email[E-mail:]{sudip.jana@okstate.edu}
\affiliation{Harish-Chandra Research Institute,  Chhatnag Road, Jhunsi, Prayagraj 211019, India}
\affiliation{Homi Bhabha National Institute, Training School Complex, Anushakti Nagar, Mumbai 400 094, India}

\author{\bf Pragyanprasu Swain}
\email[E-mail:]{pragyanprasuswain@hri.res.in}
\affiliation{Harish-Chandra Research Institute,  Chhatnag Road, Jhunsi, Prayagraj 211019, India}
\begin{abstract}
Neutrino oscillations provide compelling evidence for physics beyond the Standard Model, while the weakly interacting nature of neutrinos makes them powerful probes of new interactions and hidden sectors. In this work, we investigate a \textit{dark neutrino portal} scenario in which neutrino mass generation is linked to a light dark sector charged under a new $U(1)_D$ gauge symmetry. While Standard Model fields remain neutral under $U(1)_D$, the dark neutrino sector is charged and communicates with the Standard Model exclusively through active--dark neutrino mixing. The associated neutrinophilic mediator induces ultra-long-range interactions, whereby electrons and neutrons in the Earth, Moon, Sun, Milky Way, and the cosmological matter distribution generate sizable matter potentials that modify neutrino oscillations.  We explore the sensitivity of the upcoming Deep Underground Neutrino Experiment (DUNE), whose long baseline and pronounced matter effects make it uniquely suited to probe such interactions. We show that DUNE can access previously unexplored regions of parameter space and demonstrate that the same underlying coupling can simultaneously give rise to sizable neutrino self-interactions, including regions relevant for alleviating the Hubble tension, while remaining consistent with current neutrino oscillation constraints.
\noindent 
\end{abstract}
\maketitle
\textbf{\emph{Introduction}.--} \label{sec:I}
Following the discovery of neutrino oscillations, a wealth of experimental data has enabled increasingly precise determination of neutrino oscillation parameters~\cite{deSalas:2020pgw, NuFIT, Capozzi:2021fjo}, establishing a robust framework for describing neutrino propagation over terrestrial and astrophysical baselines. Despite this success, the origin of neutrino masses and mixing remains unknown, motivating the exploration of physics beyond the Standard Model (SM), particularly in the form of nonstandard neutrino interactions (NSIs). First introduced by Wolfenstein in 1978 \cite{Wolfenstein:1977ue}, NSIs have since attracted considerable attention as a promising avenue for probing new physics. Such interactions modify the effective matter potential \cite{Wolfenstein:1977ue, Mikheyev:1985zog, Mikheev:1986wj} experienced by neutrinos during propagation, thereby introducing additional complexities in the determination of neutrino oscillation parameters. NSIs arise naturally in a wide class of neutrino mass models (see, for instance, Ref.~\cite{Babu:2019mfe} and references therein) and have been extensively studied in the context of neutrino oscillation experiments (see Refs.~\cite{Bhupal:2019qno} for reviews). In the limit of a heavy mediator, these interactions reduce to short-range contact interactions, which have been the primary focus of the existing literature. In contrast, ultralight mediators can induce long-range interactions (LRIs), with contributions arising from the cumulative matter distributions of astrophysical and cosmological structures~\cite{Bustamante:2018mzu}. In this article, we investigate the phenomenological consequences of such long-range interactions and explore their signatures in neutrino oscillation experiments.

\begin{figure}[h!]
\centering
\includegraphics[width=0.5\textwidth]{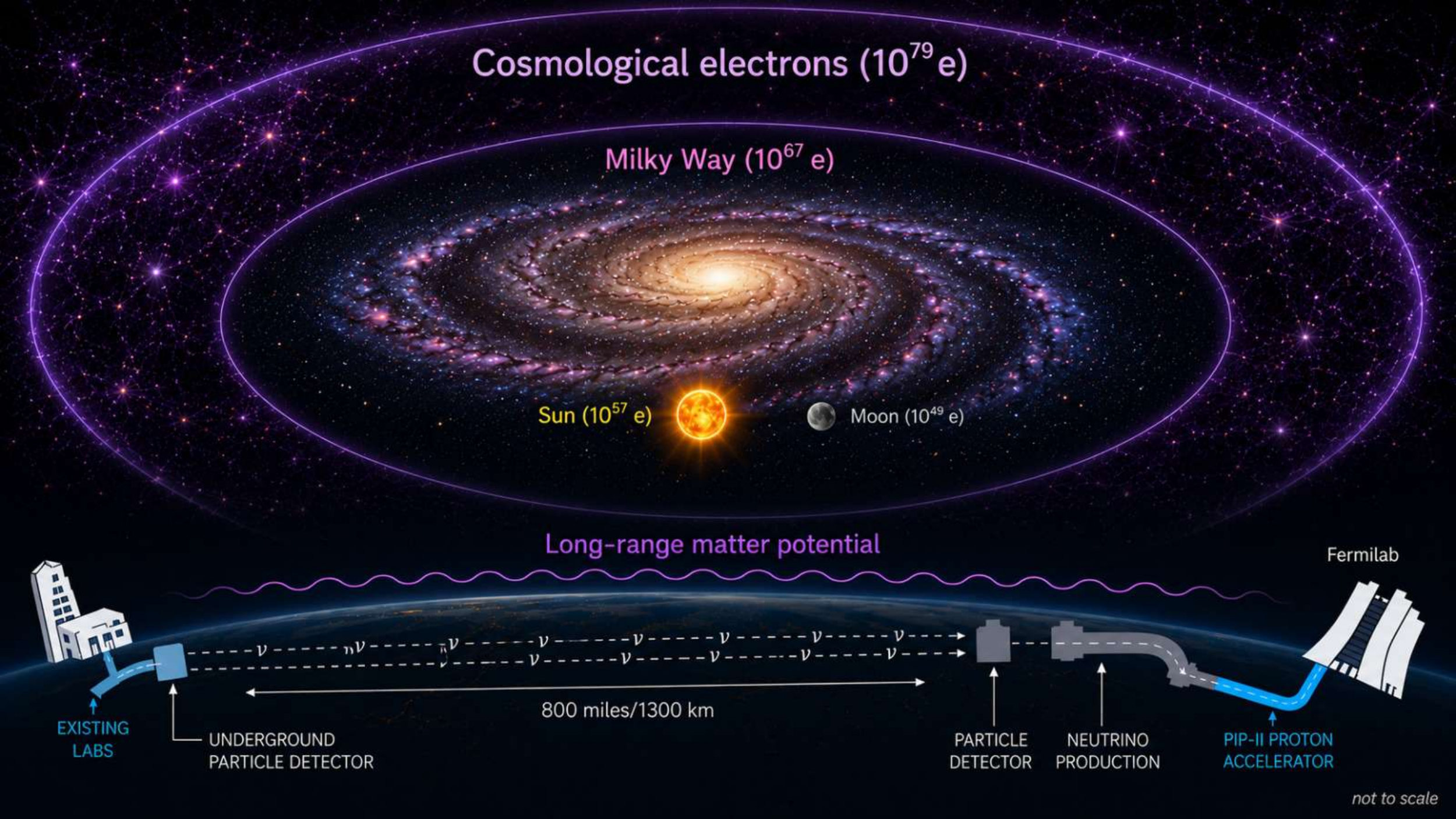}
\caption{Schematic representation of DUNE long-baseline neutrinos experiencing long-range interactions induced by the matter distribution of various celestial structures.}
\label{fig:schematic}
\end{figure}

Long-range neutrino interactions require the existence of an ultra-light mediator whose exchange generates new matter potentials affecting neutrino propagation over terrestrial and astrophysical distances. Such mediators naturally arise in Abelian gauge extensions of the SM with an additional $U(1)$ gauge symmetry.
Besides the gauged hypercharge symmetry $U(1)_Y$, the SM possesses accidental global symmetries associated with baryon and individual lepton flavor numbers, namely $U(1)_B$, $U(1)_{L_e}$, $U(1)_{L_\mu}$, and $U(1)_{L_\tau}$. Gauging anomaly-free linear combinations of these symmetries gives rise to well-motivated extensions such as $U(1)_{B-L}$, $U(1)_{L_e-L_\mu}$, $U(1)_{L_\mu-L_\tau}$, and $U(1)_{L_e-L_\tau}$. However, not all such extensions generate observable long-range neutrino interactions. In particular, flavor-universal scenarios such as $U(1)_{B-L}$ generally do not induce flavor-dependent matter effects relevant for neutrino oscillations. Furthermore, since charged leptons and quarks are also charged under these symmetries, the associated gauge bosons are subject to stringent laboratory, astrophysical, and cosmological constraints. The impact of such interactions on neutrino oscillations has been explored in long-baseline experiments such as DUNE and T2HK~\cite{Singh:2023nek, Agarwalla:2024ylc}, as well as T2HKK and P2SO~\cite{Mishra:2024riq}, and also in IceCube~\cite{Garg:2026gwx, Garg:2026rmy}, where the induced matter potential is typically diagonal in flavor space.

In contrast, we consider a dark neutrino framework based on a $U(1)_D$ gauge symmetry~\cite{Bertuzzo:2018ftf}, under which all SM fermions are neutral while the right-handed neutrinos, identified as dark neutrinos, carry non-zero charges. The new gauge boson couples directly only to the dark neutrino sector, and interactions with active neutrinos arise through active--sterile neutrino mixing. Consequently, the induced matter potential acquires a non-trivial flavor structure proportional to $U_{\alpha 4}U_{\beta 4}^{*}$, where $\alpha,\beta=e,\mu,\tau$, thereby populating the full $3\times3$ flavor-space interaction matrix and leading to rich oscillation phenomenology.
An attractive feature of this minimal framework is that it naturally realizes neutrinophilic interactions while evading stringent experimental constraints. Originally introduced to dynamically generate the small $\mu$-term of the inverse seesaw mechanism \cite{Bertuzzo:2018ftf}, it has subsequently been shown to provide a possible explanation of the long-standing MiniBooNE anomaly~\cite{Bertuzzo:2018itn, Ballett:2018ynz}, support sizable neutrino self-interactions mediated by light vector bosons \cite{Berbig:2020wve, Bally:2020yid}, and accommodate viable dark matter scenarios \cite{Abdallah:2024npf}. Such sizable neutrino self-interactions can delay the onset of neutrino free streaming in the early Universe, thereby alleviating the Hubble tension. The dark sector communicates with the SM primarily through scalar mixing, active--sterile neutrino mixing, and gauge boson kinetic or mass mixing, leading to rich phenomenological consequences across particle physics and cosmology.

In this work, we investigate the impact of such dark neutrino interactions on neutrino oscillations at the next-generation long-baseline experiment DUNE~\cite{DUNE:2020lwj, DUNE:2020jqi, DUNE:2021cuw, DUNE:2021mtg}, considering both the short- and long-range interaction regimes. 
The dark gauge boson $Z_D$ interacts strongly with dark neutrinos $\nu_D$. Via active--sterile neutrino mixing, the mediator develops effective couplings to active neutrinos, thereby inducing sizable neutrino self-interactions. Furthermore, mixing between the Standard Model $Z$ boson and $Z_D$ naturally leads to NSIs or, in the ultralight mediator regime, long-range neutrino interactions. As a consequence, large neutrino self-interactions and NSI effects are intrinsically linked and largely unavoidable in this class of models. We demonstrate that active--sterile neutrino mixing alone can induce an effective $Z$--$Z_D$ mixing, thereby generating long-range neutrino interactions with matter. In this work, we investigate the phenomenological implications of such interactions and analyze their impact on the sensitivity of DUNE.
Beyond modifying neutrino propagation in matter, the same ($Z_D$)-mediated interaction can induce sizable neutrino self-interactions~\cite{Berbig:2020wve, Bally:2020yid}, thereby establishing a connection between terrestrial oscillation experiments and cosmological observables. Such self-interactions may alleviate the Hubble tension~\cite{Kreisch:2019yzn, Park:2019ibn}. To assess the sensitivity of such interactions at DUNE, we compute oscillation probabilities and event rates in the presence of the new neutrino-matter interactions and explore the model parameter space as a function of the mediator mass and effective coupling. We then map the DUNE-sensitive regions onto the corresponding neutrino self-interaction strength, demonstrating that the same underlying interaction can give rise to a broad range of self-interaction effects. In particular, we identify regions of parameter space accessible to DUNE that lead to sizable neutrino self-interactions, including those relevant for alleviating the Hubble tension.

\vspace{0.1 in}
\textbf{\emph{Framework}.--} \label{sec:II}
In this section, we introduce the dark neutrino framework and the specific model under consideration, which forms the basis of our subsequent analysis. 

The model is based on the gauge symmetry
\[
SU(3)_C \times SU(2)_L \times U(1)_Y \times U(1)_D.
\]
The fermion sector is extended by introducing two species of right-handed neutrinos, $N$, carrying a $U(1)_D$ charge of $+1$, together with an additional species, $N'$, of charge $-1$ to ensure anomaly cancellation. All Standard Model (SM) fermions are neutral under $U(1)_D$. The scalar sector contains an $SU(2)_L$ doublet $\phi$ with dark charge $+1$, and two singlet scalars: $S_2$ with charge $+2$, whose vacuum expectation value (VEV) generates Majorana masses for the dark neutrinos through spontaneous lepton-number breaking, and $S_1$ with charge $+1$, whose VEV removes accidental global symmetries. Details of the scalar potential and mass spectrum can be found in Refs.~\cite{Bertuzzo:2018itn,Bertuzzo:2018ftf}. The Yukawa interactions relevant for neutrino mass generation are given by~\cite{Bertuzzo:2018ftf}
\begin{widetext}
\begin{equation}
\mathcal{L}_{\nu}
=
-y_{\nu}\,\overline{L}\,\tilde{\phi}\,N
+y_N\,S_2\,\overline{N}\,N^c
+y_{N'}\,S_2^{*}\,\overline{N'}\,N'^{\,c}
+m\,\overline{N'}\,N^c
+\mathrm{h.c.},
\label{eq:Lnu}
\end{equation}
\end{widetext}
where $y_{\nu}$, $y_N$, $y_{N'}$, and $m$ are flavor-space matrices.

Following the symmetry breaking sequence
\begin{widetext}
\begin{equation}
SU(2)_L \times U(1)_Y \times U(1)_D
\xrightarrow{\langle \phi \rangle = v_\phi}
U(1)_{\rm em} \times U(1)_D
\xrightarrow{\langle S_2 \rangle = \omega_2}
U(1)_{\rm em},
\end{equation}
the neutrino mass matrix in the $(\nu,\,N,\,N')$ basis becomes \cite{Bertuzzo:2018ftf}
\begin{equation}
\mathcal{M}_{\nu}
=
\frac{1}{\sqrt{2}}
\begin{pmatrix}
0 & y_{\nu}\,v_{\phi} & 0 \\
y_{\nu}^{T}\,v_{\phi} & y_N\,\omega_2 & \sqrt{2}\,m \\
0 & \sqrt{2}\,m^{T} & y_{N'}\,\omega_2
\end{pmatrix}.
\label{eq:Mnu}
\end{equation}
\end{widetext}

In the limit of small lepton-number violation, the model realizes the inverse seesaw mechanism, yielding the effective active neutrino mass matrix
\begin{equation}
m_{\nu}
\simeq
\left(y_{\nu}^{T}v_{\phi}\right)
\frac{1}{m^{T}}
\left(y_{N'}\omega_{2}\right)
\frac{1}{m}
\left(y_{\nu}v_{\phi}\right).
\label{eq:mnu_eff}
\end{equation}

The interactions of the dark gauge boson relevant for our analysis are described by the Lagrangian~\cite{Bertuzzo:2018itn,Bertuzzo:2018ftf}
\begin{widetext}
\begin{equation}
{\cal L}_{\cal D} \supset \frac{m^2_{Z_{\cal D}}}{2} \,
Z_{{\cal D}\mu} Z_{\cal D}^{\mu}
+ g_{\cal D} Z_{\cal D}^\mu \,
\overline{\nu}_{\cal D} \gamma_\mu \nu_{\cal D}
+ e \epsilon \, Z_{\cal D}^\mu \,
J_\mu^{\rm em}
+ \frac{g}{c_W} \epsilon' \, Z_{\cal D}^\mu \,
J_\mu^{Z} \, ,
\label{equ:dark_lag}
\end{equation}
\end{widetext}
where $\epsilon$ denotes the kinetic mixing between the hypercharge gauge bosons $B_{\mu\nu}$ and $B'_{\mu\nu}$, while $\epsilon'$ parametrizes the mass mixing between the SM $Z$ boson and the dark gauge boson $Z_{\cal D}$. The latter is approximately given by
\begin{equation}
\epsilon' \simeq \frac{2 g_D}{g/c_W}\,\frac{v_\phi^2}{v^2}.
\end{equation}
While the main analysis of this work assumes kinetic and mass mixing between $Z$ and $Z_D$ at the tree level, these mixings may also be generated radiatively through loop effects. This possibility is discussed in the Appendix.

The dark sector communicates with the SM through dark portals, namely neutrino mixing and kinetic as well as mass mixing in the gauge sector, leading to a rich phenomenology. In particular, the mixing between active and dark neutrinos can be expressed as
\begin{equation}
\nu_\alpha = \sum_{i=1}^{3} U_{\alpha i}\,\nu_i + U_{\alpha 4}\,N_{\cal D},
\qquad \alpha=e,\mu,\tau,{\cal D},
\label{equ:mix}
\end{equation}
where $N_{\cal D}$ denotes the dark neutrino state.
\vspace{0.1 in}

\begin{figure}[h!]
\centering
\includegraphics[width=0.4\textwidth]{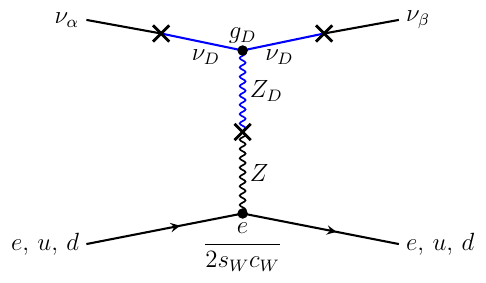}
\caption{Feynman diagram depicting the interaction between neutrinos and matter fields at the tree level.\label{fig:U1(D)-interaction}}
\end{figure}

\textbf{\emph{Matter potential and self-interactions of neutrinos}.--} \label{sec:III}
Taking into account the complete dark sector Lagrangian, active--dark neutrino mixing induces new interactions between neutrinos and matter fields. The corresponding interactions generated by Eq.~\ref{equ:dark_lag} are schematically illustrated in Fig.~\ref{fig:U1(D)-interaction}.
The dark neutrino framework induces an additional matter potential given by
\begin{align}
V_{\alpha\beta}
&=
\frac{g_D}{m^2_{Z_D}}
\frac{4 c_W m_Z^2}{g}
\,\epsilon' U_{\alpha 4}^* U_{\beta 4}
(-V_{\rm NC})
\nonumber\\
&=
\frac{g_{\rm eff}^2}{m_{Z_D}^2}
\frac{4 c_W m_Z^2}{g}
(-V_{\rm NC}),
\label{equ:NSI_pot}
\end{align}
where the effective coupling is defined as
\begin{equation}
g_{\rm eff}^2
=
g_D \epsilon' U_{\alpha 4}^* U_{\beta 4}.
\label{equ:eff_coup}
\end{equation}
The SM neutral-current matter potential is
\begin{equation}
V_{\rm NC}
=
-\frac{1}{2}\sqrt{2}\,G_F n_n,
\label{equ:mat-potential-NC}
\end{equation}
with $G_F$ denoting the Fermi constant and $n_n$ the neutron number density. The new matter potential arises solely from the $Z$--$Z_D$ mass mixing term, while the kinetic mixing contribution vanishes in electrically neutral, static, and unpolarized matter.

As evident from Eq.~\ref{equ:mat-potential-NC}, the framework under consideration generally gives rise to a fully populated $3\times3$ interaction potential matrix in flavor space. To facilitate a transparent interpretation of the underlying flavor structure, we consider three benchmark textures in which only one $2\times2$ submatrix of the full interaction potential is nonzero at a time. This approach enables us to isolate the dominant flavor effects while avoiding intricate correlations inherent to the general three-flavor scenario. The benchmark configurations are defined as follows:

\begin{enumerate}
    \item $|U_{e4}| = |U_{\mu4}|$, $|U_{\tau4}|=0$, corresponding to a scenario in which only the $e\mu$ sector of the new interaction potential matrix is populated.
    
    \item $|U_{e4}| = |U_{\tau4}|$, $|U_{\mu4}|=0$, corresponding to a scenario in which only the $e\tau$ sector of the new interaction potential matrix is populated.
    
    \item $|U_{\mu4}| = |U_{\tau4}|$, $|U_{e4}|=0$, corresponding to a scenario in which only the $\mu\tau$ sector of the new interaction potential matrix is populated.
\end{enumerate}

To assess the sensitivity of our analysis to the dark-neutrino framework, we investigate both the short- and long-range regimes of neutrino--matter interactions, thereby providing a unified probe over a wide range of mediator masses. For mediator masses satisfying $m_{Z_D} \gtrsim 10^{-12}$~eV, where the interaction range is significantly shorter than the neutrino oscillation length, the new interaction effectively reduces to a contact potential proportional to $1/m_{Z_D}^2$, as given in Eq.~\ref{equ:mat-potential-NC}. This regime is commonly referred to as the short-range or non-standard interaction (NSI) regime~\cite{Coloma:2020gfv}. In contrast, mediator masses in the range $m_{Z_D} \lesssim 10^{-12}$~eV correspond to the long-range interaction regime.

The long-range interaction (LRI) is described by the Yukawa potential~\cite{Bustamante:2018mzu, Agarwalla:2023sng, Singh:2023nek, Agarwalla:2024ylc},
\begin{equation}
 V_{{\rm LRI}, f}
 =
 g^2_{\rm eff}
 \frac{e^{-m_{Z_D}d}}{4\pi d}
  N_f\;,
 \label{equ:Yukawa_potential_zD}
\end{equation}
where $d$ denotes the distance between the neutrino and the source, and $N_f$ represents the number of fermions (electrons, protons, or neutrons) in the matter distribution sourcing the potential. In the dark-gauge boson mediated neutrino--matter interaction considered here (see Fig.~\ref{fig:U1(D)-interaction}), only the neutron contribution survives, such that $N_f = N_n$.

For ultralight mediators with masses in the range $10^{-35}$--$10^{-10}$~eV, the interaction range extends from kilometer to gigaparsec scales. Consequently, the total long-range potential receives contributions from the Earth ($\oplus$), Moon ($\leftmoon$), Sun ($\odot$), Milky Way (MW), and the cosmological matter distribution (cos). The total potential can therefore be written as
\begin{widetext}
\begin{equation}
 \label{equ:pot_tot}
 V_{{\rm LRI}, n}
 (m_{Z_D}, g_{\rm eff})
 = 
 V_{{\rm LRI}, n}^\oplus
 +
 V_{{\rm LRI}, n}^{\leftmoon}
 +
 V_{{\rm LRI}, n}^{\astrosun}
 +
 V_{{\rm LRI}, n}^{\rm MW}
 + 
 V_{{\rm LRI}, n}^{\rm cos} 
 \;.
\end{equation}
\end{widetext}

Throughout this work, we assume that the matter sourcing the potential is electrically neutral, i.e., it contains equal numbers of electrons and protons ($N_e = N_p$), and is approximately isoscalar, i.e., it contains equal numbers of protons and neutrons ($N_p = N_n$), except for the Sun~\cite{Heeck:2010pg} and the cosmological matter distribution~\cite{Hogg:1999ad, Steigman:2007xt, Planck:2015fie}.

We treat the Sun ($N_{e,\odot} = N_{p,\odot} \sim 10^{57}$, $N_{n,\odot} \simeq N_{e,\odot}/4$) and the Moon ($N_{e,\leftmoon} = N_{p,\leftmoon} = N_{n,\leftmoon} \sim 5 \times 10^{49}$) as point sources of electrons, protons, and neutrons. In contrast, the Earth ($N_{e,\oplus} = N_{p,\oplus} = N_{n,\oplus} \sim 4 \times 10^{51}$), the Milky Way ($N_{e,\rm MW} = N_{p,\rm MW} \simeq N_{n,\rm MW} \sim 10^{67}$), and the cosmological matter distribution ($N_{e,\rm cos} = N_{p,\rm cos} \sim 10^{79}$, $N_{n,\rm cos} \sim 10^{78}$) are modeled as continuous matter distributions. The corresponding potentials from the Earth, Milky Way, and cosmological matter are computed following the methodology of Refs.~\cite{Bustamante:2018mzu, Agarwalla:2023sng, Singh:2023nek, Agarwalla:2024ylc}. While deriving the total potential due to Earth, we compute the average potential at the point of neutrino detection. This provides an effective description of Earth's contribution to the total potential. A more detailed treatment that accounts for the modification of the potential along the neutrino path inside Earth can be found in Refs.~\cite{Wise:2018rnb, Coloma:2020gfv}.

\begin{figure}[h!]
\centering
\includegraphics[width=0.47\textwidth]{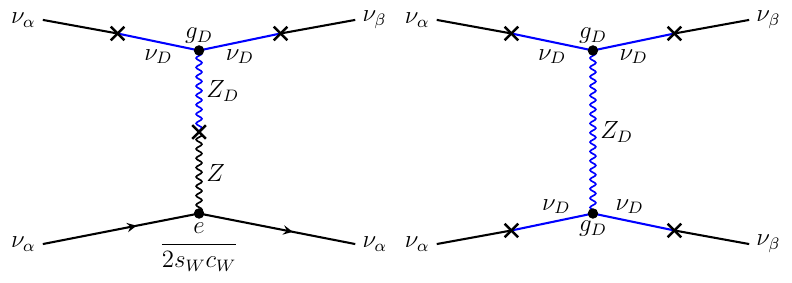}
\caption{The Feynman diagram representing the neutrino self-interactions.\label{fig:Nu-self-interaction}}
\end{figure}

An important consequence of the framework under consideration is that the same interaction responsible for neutrino--matter effects also gives rise to non-standard neutrino self-interactions (NSSI). Owing to the mixing between active and dark-sector neutrinos, the light neutrino states inherit couplings to the dark gauge boson, which mediates neutrino self-interactions. Notably, the same mixing responsible for generating the effective neutrino matter potential also governs these self-interactions. As a result, neutrino oscillation effects in matter and neutrino self-interactions are intrinsically linked and arise from a common set of underlying parameters.
The Lagrangian can be expressed as:
\begin{equation}
    \mathcal{L_{\rm NSSI}}=G^{4\nu}_{\rm eff} ({\bar\nu}\nu)({\bar\nu}\nu)\,,
\end{equation}
where,
\begin{eqnarray}
        G^{4\nu}_{\rm eff} = &\frac{1}{m^2_{Z_D}}({g^2_D |U_{\alpha 4}|^2|U_{\beta 4}|^2+g_D (g/c_W) \epsilon^\prime |{U_{\alpha 4}}| |{U_{\beta 4}|}})\nonumber\\
        =&\frac{g_D|U_{\alpha 4}||U_{\beta 4}|}{m^2_{Z_D}} (g_D|U_{\alpha 4}||U_{\beta 4}|+\epsilon^\prime (g/c_W)).
        \label{equ:self-int-coupling}
\end{eqnarray}
The dark-neutrino-induced neutrino self-interaction processes are illustrated in Fig.~\ref{fig:Nu-self-interaction}.
The model unifies neutrino matter effects and self-interactions through a common set of parameters, establishing a direct connection between neutrino oscillation phenomenology and cosmological observable. In particular, sizable neutrino self-interactions mediated by light vector bosons may delay neutrino free streaming in the early Universe, potentially alleviating the Hubble tension~\cite{Berbig:2020wve, Bally:2020yid}.

\vspace{0.1 in}
\textbf{\emph{Experimental details and Statistical methods}.--}\label{sec:IV}
To forecast the sensitivity to the model parameters, we consider the forthcoming long-baseline neutrino oscillation experiment, DUNE~\cite{DUNE:2020lwj, DUNE:2020jqi, DUNE:2021mtg, DUNE:2021cuw}. The experiment will employ a 40 kt liquid-argon time-projection chamber (LArTPC) as the far detector, situated 1285 km from Fermilab at the Homestake Mine. DUNE will utilize an on-axis wide-band neutrino beam generated by 120 GeV protons with a beam power of 1.2 MW, corresponding to an annual exposure of $1.1\times10^{21}$ protons-on-target. The neutrino energy spectrum spans approximately 0.5--10 GeV and peaks near 2.5 GeV, close to the first oscillation maximum for this baseline.

In our simulations, we adopt the configuration described in \Refe~\cite{DUNE:2021cuw}. DUNE is assumed to operate for a total of 10 years, equally divided in neutrino and antineutrino modes. To provide conservative projections, we consider only the completed DUNE configuration and neglect the smaller exposure expected during the staged construction phase~\cite{DUNE:2021tad}. Events are binned in reconstructed neutrino energy as follows: uniform bins of width 0.125 GeV in the range 0.5--8 GeV, 1 GeV bins between 8--10 GeV, and 2 GeV bins from 10--18 GeV.

To derive constraints on the Dark neutrino parameter space, we consider the Poissonian $\chi^2$ having the following form
\begin{widetext}
\begin{equation}
	\chi^{2}
	(\vec{\lambda}, \boldsymbol{\theta})
	=
	\underset{\left\{\xi_{s}, \xi_{b}\right\}}{\mathrm{min}} 
	\left\{
	2\sum^{N_e}_{i=1}
	\left[
	N_{e, i}^{{\rm test}}
	(\vec\lambda, \boldsymbol{\theta}, \xi_s, \xi_{b})
	\right. \right.
	\nonumber \\
	\left. \left.
	-
	N_{e, i}^{{\rm true}}
	\left( 
	1
	+
	\ln
	\frac{N_{e, i}^{{\rm test}}
		(\vec\lambda, \boldsymbol{\theta}, \xi_s, 
		\xi_{b})}
	{N_{e, i}^{{\rm true}}}
	\right)
	\right]
	+
	\xi^{2}_{s}
	+
	\xi^{2}_{b} 
	\right\} \;, \;\,\,\,\,\,\,\,\,\,
	\label{equ:chi2_definition}
\end{equation}
\end{widetext}
where $\boldsymbol{\theta} \equiv \{\theta_{23}, \delta_{\rm CP}\}$ are the most uncertain neutrino oscillation parameters which we vary in the test hypothesis in their allowed $3\sigma$ ranges while keeping others fixed to their best-fit (true) values from NuFIT 5.1~\cite{Esteban:2020cvm, NuFIT} with normal neutrino mass ordering without Super-K atmospheric data. $\vec\lambda$ contains the new physics parameters. The test events ($N_{e, i}^{{\rm test}}$) and true events ($N_{e, i}^{{\rm true}}$) are generated with $\vec\lambda\neq0$ and $\vec\lambda=0$, respectively. Both the true and test event spectra are calculated by summing over the signal and all the backgrounds. $\xi_{s}$ and $\xi_{b}$ are the systematic pulls over the signal and background, respectively. See \Refe~\cite{Agarwalla:2024ylc} for details of the backgrounds and systematics associated with the signals and backgrounds (adopted from DUNE TDR~\cite{DUNE:2021cuw}). The neutrino mass ordering is assumed to be normal while generating true and test events.

Our results are presented in terms of the following test statistic
\begin{equation}
	\label{equ:delta_chi2}
	\Delta\chi^2 (\vec{\lambda})
	=
	\underset
	{\{ \boldsymbol{\theta}\}}{\mathrm{min}} 
	\left[
	\chi^{2}
	(\vec\lambda, \boldsymbol{\theta}) 
	-
	\chi_{{\rm min}}^{2}
	\right]
	\;,
\end{equation}

where $\chi_{{\rm min}}^{2}$ is calculated for $\boldsymbol\theta=\boldsymbol{\theta}^{\rm true}$ and $\vec{\lambda}=0$.

\vspace{0.1 in}
\textbf{\emph{Oscillation probability and event spectra in DUNE}.--}
\label{sec:V}
\begin{figure*}[h!]
\centering
\includegraphics[width=\textwidth]{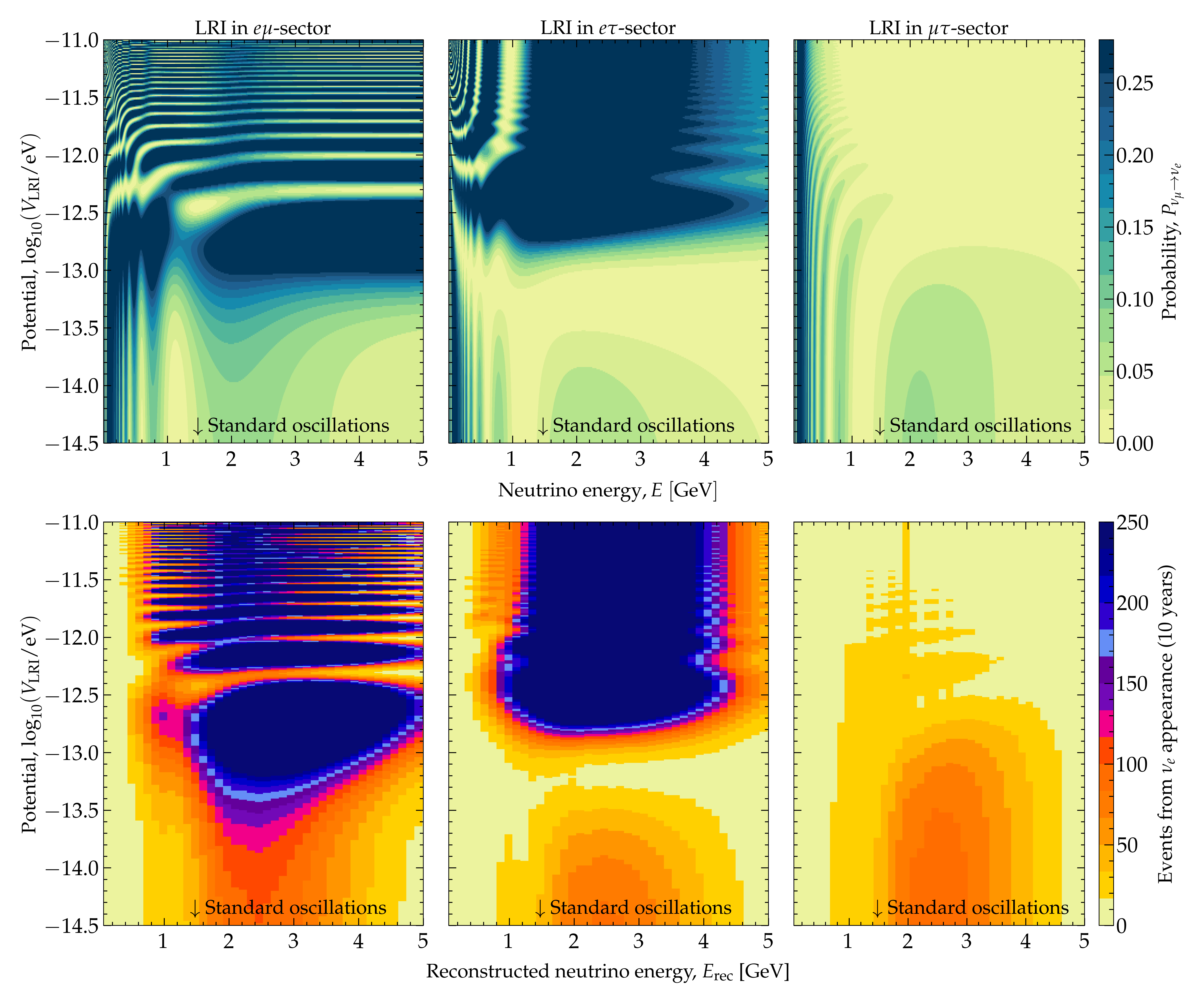}
\caption{Neutrino oscillation probabilities and event spectra at DUNE in the presence of the dark gauge boson-induced interaction described in Eq.~\ref{equ:dark_lag}. The left, middle, and right panels show the cases where the corresponding interaction potential arises in the $e\mu$, $e\tau$, and $\mu\tau$ sectors, respectively.
}
\label{fig:prob_event}
\end{figure*}
The propagation Hamiltonian of a neutrino in matter is given by
\begin{equation}
 \label{equ:hamiltonian_tot}
 \mathbf{H}
 =
 \mathbf{H}_{\rm vac}
 +
 \mathbf{V}_{\rm mat}
 +
 \mathbf{V}_{\rm new} \;.
\end{equation}
where
\begin{equation}
 \label{equ:hamiltonian_vac}
 \mathbf{H}_{\rm vac}
 =
 \frac{1}{2 E}
 \mathbf{U_{\rm PMNS}}~
 {\rm diag}(0, \Delta m^2_{21}, \Delta m^2_{31})
 ~\mathbf{U^{\dagger}_{\rm PMNS}} \;,
\end{equation}\,
governs the oscillations in vacuum,
\begin{equation}
 \label{equ:v_mat}
 \mathbf{V}_{\rm mat}
 =
 {\rm diag}(V_{\rm CC}, 0, 0) \;,
\end{equation}
is the standard-matter potential, with $V_{\rm CC} = \sqrt{2} G_F n_e$, $G_F$ denoting the Fermi constant, and $n_e$ the electron number density.

The new neutrino-matter potential, $\mathbf{V}_{\rm new}$, induced by the dark gauge boson $Z_D$, is given by Eq.~\ref{equ:NSI_pot} for short-range interactions. For the long-range regime, however, it has the form of the Yukawa potential, Eq.~\ref{equ:Yukawa_potential_zD}.
We implement the new interaction potential in the probability engine $snu.c$~\cite{Kopp:2006wp, Kopp:2007ne} of the General Long Baseline Experiment Simulator~\cite{Huber:2004ka, Huber:2007ji} (GLoBES) to examine how the long-range interaction (LRI) affects neutrino oscillation probabilities and the corresponding event rates at DUNE.
The upper panels of Fig.~\ref{fig:prob_event} show the $P_{\nu_\mu\to\nu_e}$ appearance probability as a function of neutrino energy and the strength of the LRI potential. The lower panels show the corresponding event rates for 10 years of exposure. The left, middle, and right panels correspond to cases where new physics is present in the $e\mu$, $e\tau$, and $\mu\tau$ sectors of the new potential, respectively. When the strength of the new potential is small, the oscillations are driven by $\mathbf{H}_{\rm vac}+\mathbf{V}_{\rm mat}$ and are labeled as standard oscillations in all panels. As the strength of the new potential increases and the condition $ \mathbf{V}_{\rm new} \approx \mathbf{H}_{\rm vac}+\mathbf{V}_{\rm mat}$ is achieved, the atmospheric resonance occurs, and the oscillation probability attains its maximum value (darkest regions prominent in case of $e\mu$ and $e\tau$ sectors). Compared to ref.~\cite{Agarwalla:2024ylc}, where the resonance regions have blob-like structures, here these are broadened due to the presence of non-zero off-diagonal entries in the $\mathbf{V}_{\rm new}$ matrix. The resonance is not observed when new physics is present in the $\mu\tau$ sector because it effectively reduces the standard matter potential $V_{\rm CC}$, so the resonance is not reached. Hence, we expect lower sensitivity when new physics affects the $\mu\tau$ sector than the $e\mu$ and $e\tau$ sectors. Similar features are also observed in the event rates.
\vspace{0.1 in}

\textbf{\emph{Projected Sensitivity of DUNE}.--}
\label{sec:VI}
Having discussed the oscillation probabilities, event spectra, and statistical framework in the preceding sections, we now present the projected sensitivity of DUNE to the parameter space of the dark-neutrino portal model. The procedure for deriving sensitivities differs between the long-range and short-range interaction regimes. In the long-range interaction (LRI) scenario, the oscillation probabilities are affected by an effective matter potential that depends on both the mediator mass and the matter distribution within the source. We therefore first constrain the effective potential at the oscillation level and subsequently map the resulting sensitivities onto the mediator mass--coupling plane using Eq.~\ref{equ:Yukawa_potential_zD}. By contrast, in the short-range limit, the interaction effectively reduces to a contact term proportional to $g_{\rm eff}^2/m_{Z_D}^2$. In this regime, the induced matter potential can be expressed directly in terms of the model parameters through Eq.~\ref{equ:NSI_pot}, enabling a direct determination of sensitivities in the mediator mass--coupling plane without requiring an intermediate mapping procedure.

\begin{figure}[t!]
\centering
\includegraphics[width=0.5\textwidth]{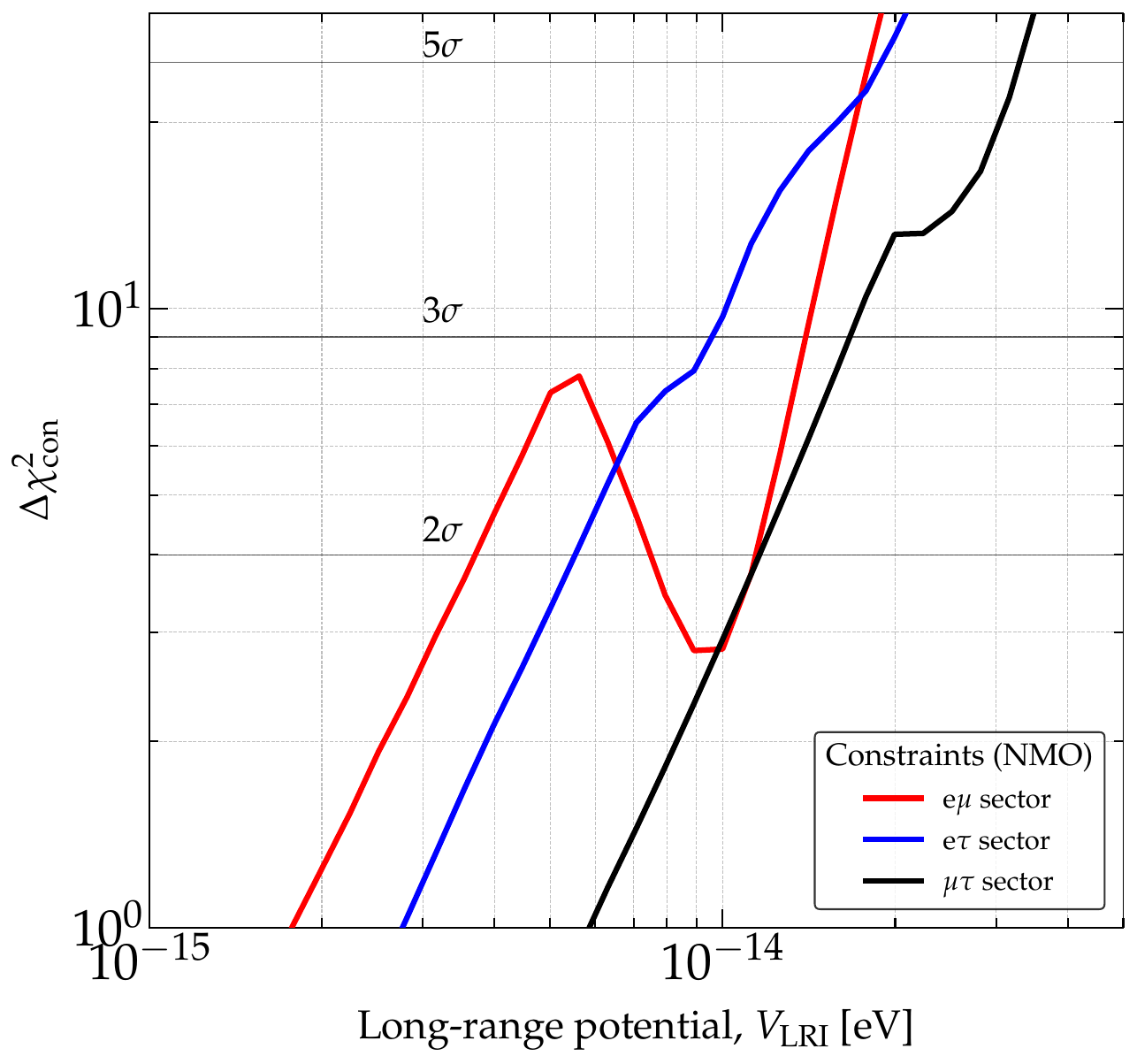}
\caption{Projected limits on the new physics potential in the $e\mu$, $e\tau$, and $\mu\tau$ sectors with ten years of exposure of DUNE. We minimize the $\Delta\chi^2$ over the uncertain oscillation parameters, $\theta_{23}$ and $\delta_{\rm CP}$, and the phase associated with the off-diagonal elements of the new interaction potential matrix. We assume normal neutrino mass ordering.}
\label{fig:LRI-potential}
\end{figure}

Fig.~\ref{fig:LRI-potential} shows the projected sensitivity to the long-range interaction potential induced by the dark-neutrino model, taking the benchmark flavor textures introduced in the previous section. The $3\sigma$ upper limits on $\mathbf{V_{\rm LRI}}$ are given in table~\ref{tab:VLRI_limits}. As expected from the effect on probability and event rates (Fig.~\ref{fig:prob_event}), the limits are tighter in the case of $e\mu$ and $e\tau$ sectors, and is the weakest in the case of $\mu\tau$ sector. 

\begin{table}[h!]
		\centering
		\begin{tabular}{c c}
			\hline\hline
			\textbf{Flavor sector} & $\mathbf{|V_{\rm LRI}| \;(\text{eV})}$ \\
			\hline
			$e\mu$   & $1.401 \times 10^{-14}$ \\
			$e\tau$  & $ 9.604 \times 10^{-15}$ \\
			$\mu\tau$ & $ 1.672 \times 10^{-14}$ \\
			\hline\hline
		\end{tabular}
\caption{$3\sigma$ upper bounds on the long-range potential $V_{\rm LRI}$ obtained from the oscillation seen at DUNE with ten years of its exposure for different flavor sectors. We translate these values to the mass-coupling plane shown in Fig.~\ref{fig:mass-coupling-plane}.}
\label{tab:VLRI_limits}
\end{table}

\begin{figure*}[h!]
\centering
\includegraphics[width=1\textwidth]{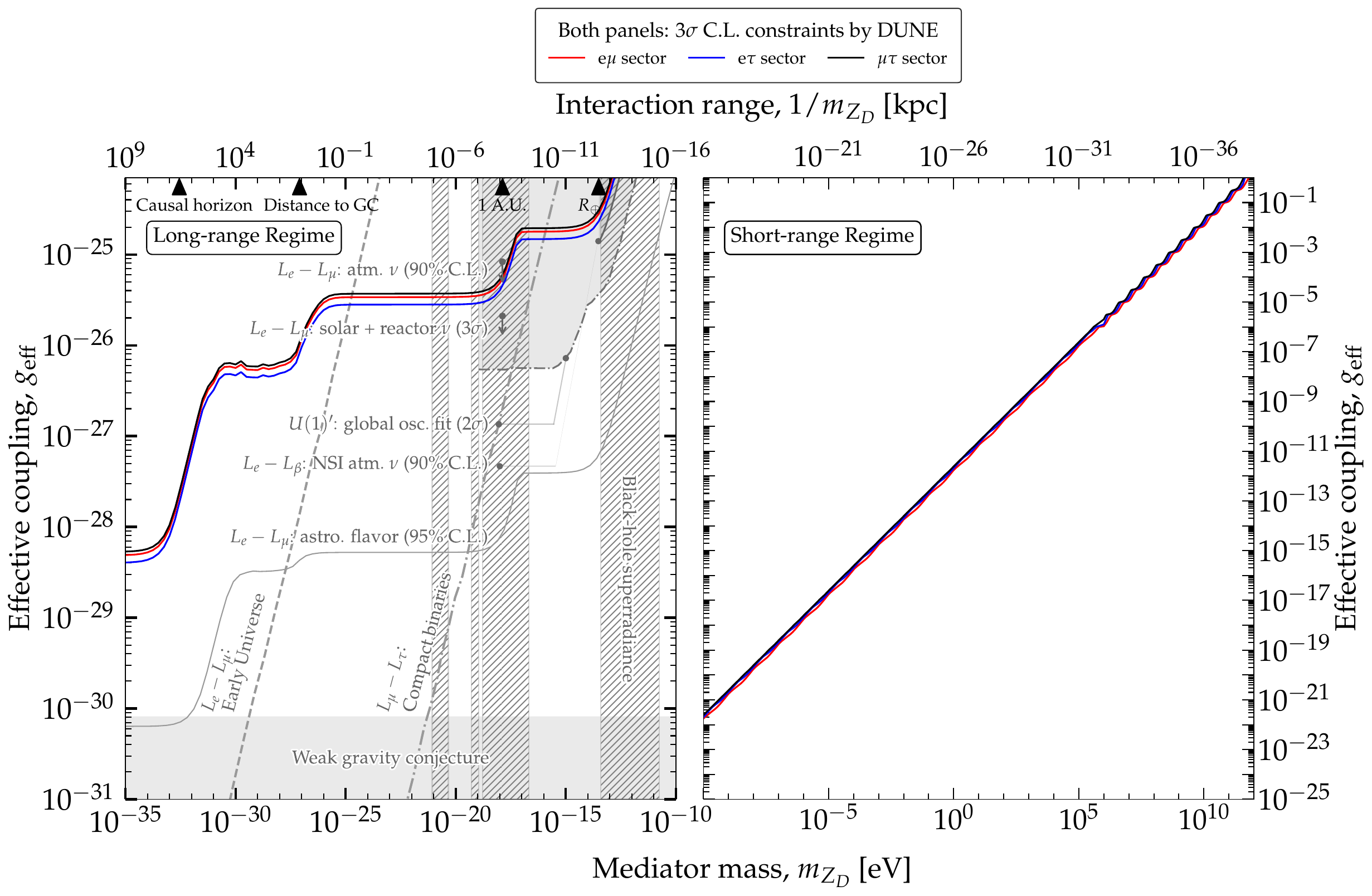}
\caption{Projected sensitivity on the mass vs coupling plane of the new light mediator from $U(1)_D$ model with ten years of exposure of DUNE. The results are for cases in which new physics is present in the $e\mu$, $e\tau$, and $\mu\tau$ sectors. The limits are translated from $3\sigma$ limits on the potential in Fig.~\ref{fig:LRI-potential}. We assume normal neutrino mass ordering. Existing constraints shown in the left panel are primarily obtained from a recent global oscillation analysis~\cite{Coloma:2020gfv}, where the limit shown is the lower envelope across other $U(1)^\prime$ symmetries considered. Additional bounds arise from atmospheric neutrinos~\cite{Joshipura:2003jh}, solar and reactor neutrino data~\cite{Bandyopadhyay:2006uh}, and studies of non-standard neutrino interactions~\cite{Super-Kamiokande:2011dam, Ohlsson:2012kf, Gonzalez-Garcia:2013usa}. The projected sensitivity, based on the current flavor-composition measurement of high-energy astrophysical neutrinos at IceCube, is taken from Ref.~\cite{Agarwalla:2023sng}. Indirect constraints arise from Black-hole Superradiance~\cite{Baryakhtar:2017ngi, Davoudiasl:2019nlo}, Weak gravity conjecture~\cite{Arkani-Hamed:2006emk} with the lightest neutrino mass of $0.01$~eV, Early Universe~\cite{Dror:2020fbh}, and Compact binary systems~\cite{KumarPoddar:2019ceq}.}
\label{fig:mass-coupling-plane}
\end{figure*}

\begin{figure*}[t!]
\centering
\includegraphics[width=0.8\textwidth]{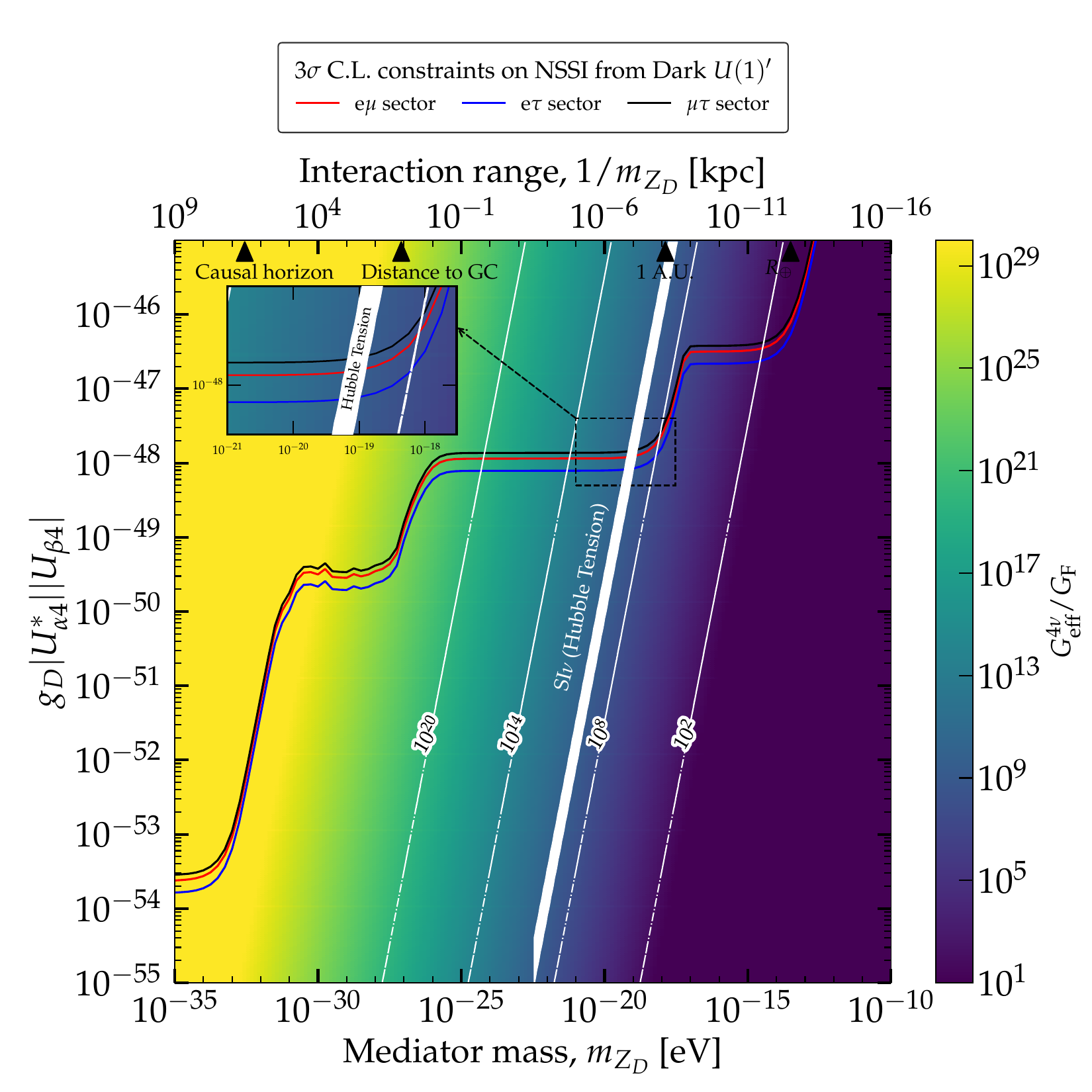}
\caption{Projected $3\sigma$ sensitivity of DUNE to the effective non-standard neutrino self-interaction (NSSI) strength, $G^{4\nu}_{\rm eff}/G_F$, as a function of the mediator mass $m_{Z_D}$, inferred from the dark-neutrino-induced effective coupling $g_D |U^{*}_{\alpha 4}||U_{\beta 4}|$ ($= g_{\rm eff}/\epsilon^\prime$, with $\epsilon^\prime = 10^{-3}$). The colored curves show the projected DUNE sensitivities for long-range interactions in the $e\mu$, $e\tau$, and $\mu\tau$ sectors. The white contours correspond to constant values of $G^{4\nu}_{\rm eff}/G_F$, while the shaded white band denotes the region favored for resolving the Hubble tension~\cite{Kreisch:2019yzn, Blinov:2019gcj}.}
\label{fig:self-int-hubble-tension}
\end{figure*}

Fig.~\ref{fig:mass-coupling-plane} shows the $3\sigma$ upper limits on the plane of mass and coupling of the dark gauge boson $Z_D$. The left panel corresponds to the long-range regime. The limits on the potential (refer to table~\ref{tab:VLRI_limits}) are translated to this plane. The step-like structures in each curve are due to the addition of a new celestial object at the transition point, contributing to the total potential, thus enhancing the sensitivity~\cite{Bustamante:2018mzu, Agarwalla:2023sng, Singh:2023nek, Agarwalla:2024ylc}. Our limits are competitive with existing constraints on the long-range interaction of neutrinos, as shown by the gray lines and bands. The right panel shows the constraints in the short-range regime. In this regime, the sensitivity follows the expected scaling with $g^2_{\rm eff}/m^2_{Z_D}$, leading to a straight line nature in the mass-coupling plane.

As discussed earlier, the same interactions responsible for neutrino--matter effects in the dark-neutrino portal framework can also induce non-standard neutrino self-interactions (NSSI). In this scenario, the dark sector communicates with the Standard Model through both neutrino mixing and gauge-boson kinetic or mass mixing portals. The mixing between active and dark neutrinos generates direct couplings of the dark gauge boson, $Z_D$, to active neutrinos, thereby inducing neutrino self-interactions. At the same time, the mass mixing between the $Z$ and $Z_D$ bosons gives rise to neutrino--matter interactions, including long-range forces. Consequently, the same active--dark neutrino mixing parameters govern both phenomena. For a fixed value of the gauge-boson mixing parameter $\epsilon^\prime$, the neutrino mixing parameters simultaneously determine the neutrino matter potential and the strength of neutrino self-interactions.
To illustrate this intrinsic connection within the dark-neutrino portal model, in Fig.~\ref{fig:self-int-hubble-tension} we recast the oscillation sensitivities into the effective neutrino self-interaction strength normalized to the Fermi constant, $G^{4\nu}_{\rm eff}/G_F$, using Eq.~\ref{equ:self-int-coupling}. The background color gradient represents the value of $G^{4\nu}_{\rm eff}/G_F$ implied by the dark-neutrino-induced effective coupling. The white contours correspond to constant values of $G^{4\nu}_{\rm eff}/G_F$, while the shaded white band indicates the $3\sigma$ region favored for alleviating the Hubble tension, corresponding to $G^{4\nu}_{\rm eff}=(4.7^{+1.2}_{-1.8},{\rm MeV})^{-2}$~\cite{Kreisch:2019yzn, Blinov:2019gcj}. The colored curves denote the oscillation sensitivities in different flavor sectors. In particular, light mediators can generate extremely large self-interaction strengths even for relatively small couplings, whereas the interaction strength becomes suppressed for heavier mediators. Interestingly, part of the parameter space accessible to oscillation experiments also overlaps with regions yielding cosmologically relevant self-interactions capable of alleviating the Hubble tension.

\vspace{0.1 in}
\textbf{\emph{Conclusions}.--}
\label{sec:VII}
In this work, we investigated dark neutrino-induced non-standard neutrino-matter interactions in both short- and long-range regimes, together with their interplay with neutrino self-interactions, at DUNE. Using a 10-year exposure, we derived sensitivities to the effective dark neutrino coupling across mediator masses spanning $10^{-35}\!-\!10^{12}\,\mathrm{eV}$. Adopting benchmark textures with a single nonzero $2\times2$ flavor block, we showed that the oscillation phenomenology is strongly flavor dependent. While the $e\mu$ and $e\tau$ sectors exhibit resonance-like features in oscillation probabilities and event rates, the $\mu\tau$ sector does not, leading to distinct sensitivities. In particular, DUNE can probe effective couplings as small as $g_{\rm eff}\lesssim10^{-28}$ for mediator masses of order $10^{-35}\,\mathrm{eV}$.
We translated the oscillation sensitivity into the corresponding regions in the mediator mass--effective coupling parameter space, and mapped them onto the corresponding neutrino self-interaction strength. We find that values of the four-neutrino self-interaction strength relevant for alleviating the Hubble tension, $G_{\rm eff}^{4\nu}\sim10^{9}\!-\!10^{10}\,G_{\rm F}$, remain compatible with current oscillation data, while a portion of the parameter space can be probed by DUNE.
Overall, our results demonstrate that searches for dark neutrino-induced long-range neutrino-matter interactions at DUNE provide a novel probe of neutrino mass generation in secluded dark sectors, while simultaneously exploring regions of parameter space that realize cosmologically relevant neutrino self-interactions.

\vspace{0.1 in}
\textbf{\emph{Acknowledgments}.--}
The authors would like to thank Santosh Kumar Rai for useful discussions. SJ and PS acknowledge the support from the Department of Atomic Energy (DAE), Government of India. The simulation work has been carried out using the High Performance Cluster Facility provided by the Harish-Chandra Research Institute. The results and conclusions presented in this work are the sole opinions of the authors and do not represent the official DUNE collaboration. 

\vspace{0.1 in}
\begin{widetext}
\section*{Appendix}
\textbf{\emph{Loop-induced neutrino-matter interactions}.--}
\begin{figure}[t!]
\centering
\includegraphics[width=0.4\textwidth]{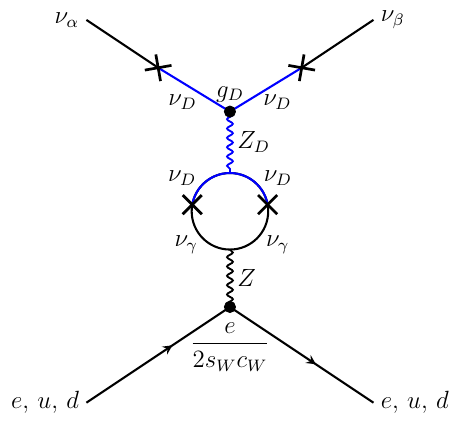}
\includegraphics[width=0.4\textwidth]{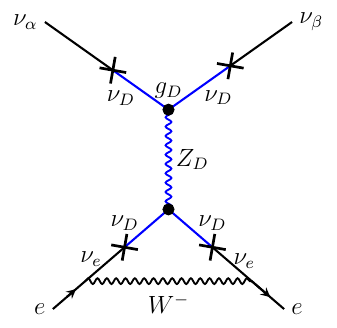}
\caption{The Feynman diagram showing the loop-induced neutrino-matter interactions.\label{fig:U1(D)-interaction-loop}}
\end{figure} 
\begin{figure}[t!]
\centering
\includegraphics[width=0.6\textwidth]{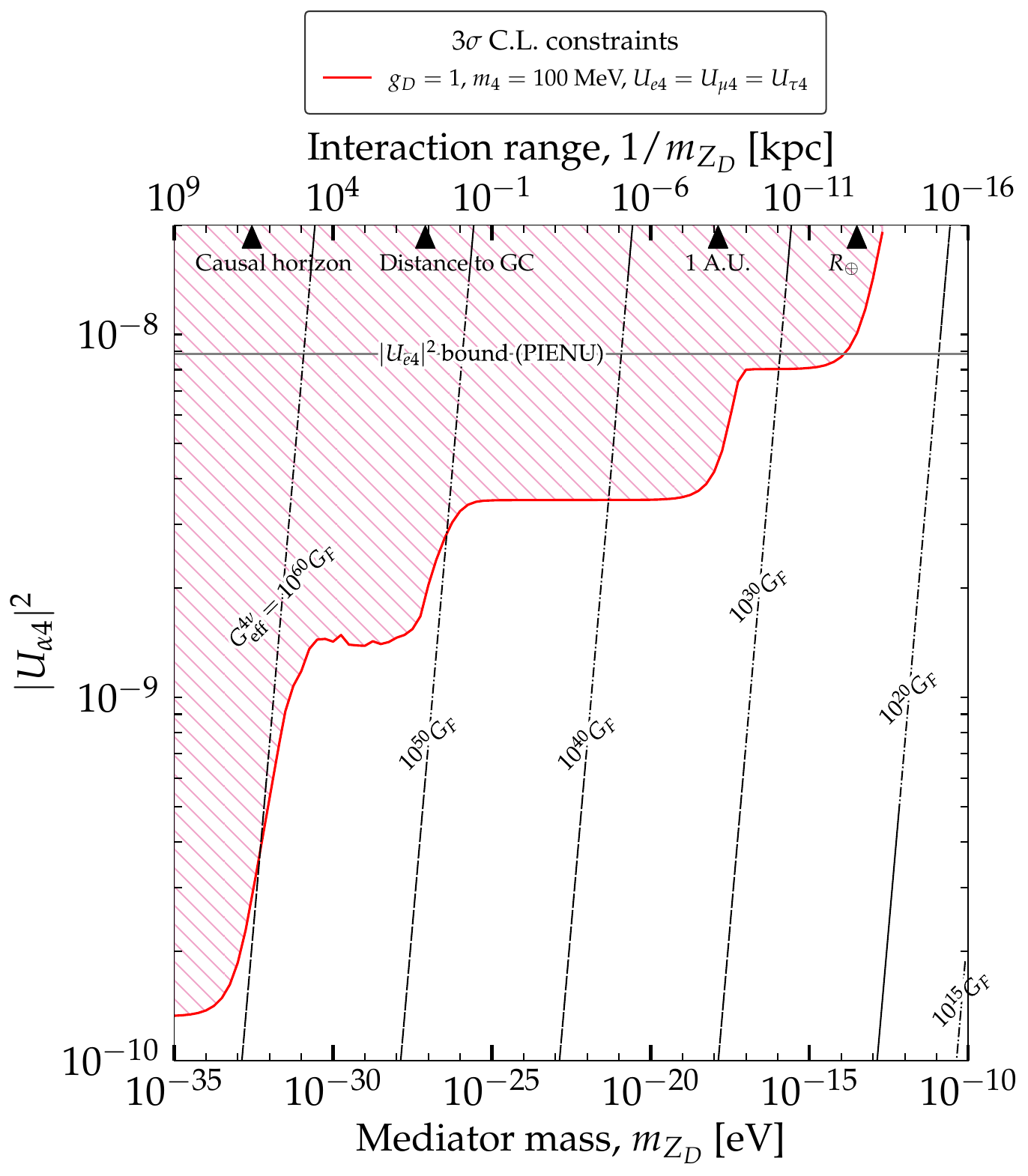}
\caption{Translated upper limits (derived from Fig.~\ref{fig:mass-coupling-plane}) on the active-dark neutrino mixing, where the $Z-Z_D$ mixing is induced via loop (see Fig.~\ref{fig:U1(D)-interaction-loop}). The shaded regions are ruled out at $3\sigma$ C.L. The strongest existing limit on $|U_{e 4}|^2$ is from PIENU~\cite{Bryman:2019ssi, Bryman:2019bjg} which is $\lesssim 8.838 \times 10^{-9}$. The strongest limits on $|U_{\mu 4}|^2$ and $|U_{\tau 4}|^2$, obtained from KEK~\cite{Hayano:1982wu}  and CHARM~\cite{Orloff:2002de}, respectively, are not shown in this figure; these correspond to $\lesssim 6 \times 10^{-6}$ and $\lesssim 7.37 \times 10^{-4}$. The existing bounds are collected from~\Refe~\cite{Bolton:2019pcu}, which gives an exhaustive list of bounds.
\label{fig:loop-mixing-limit}}
\end{figure}
Notably, even in the absence of tree-level ($Z-Z_D$) mass mixing, neutrino-matter interactions can still arise through quantum corrections at the one-loop level (see Fig.~\ref{fig:U1(D)-interaction-loop}).
The effective potential is given by
\begin{equation}
V_{\alpha\beta, Z} = \frac{g^2_D}{m^2_{Z_D}} \frac{m^2_4 |U_{\gamma 4}|^2}{8 \pi^2}U^{*}_{\alpha 4} U_{\beta 4} (-V_{\rm NC})\,.
\end{equation}

The tree-level potential in Eq.~\ref{equ:NSI_pot} can also be generated radiatively through neutrino-loop induced \(Z\text{-}Z_D\) mixing by replacing the mass-mixing parameter \(\epsilon^\prime\) with the loop-induced mixing parameter \(\epsilon_{\rm loop}\), given by
\begin{equation} 
\epsilon_{\rm loop} = \frac {g g_D}{4 c_W m^2_Z} \frac{m^2_4 |U_{\gamma 4}|^2}{8 \pi^2}\,.
\end{equation}

Likewise, when the neutrino-matter interaction is mediated by a loop involving \(W^\pm\) bosons (see Fig.~\ref{fig:U1(D)-interaction-loop}), the resulting effective potential takes the form
\begin{equation}
V_{\alpha\beta, W} = \frac{g^2_D}{m^2_{Z_D}} \frac{m^2_4 |U_{\gamma 4}|^2}{8 \pi^2}U^{*}_{\alpha 4} U_{\beta 4} (-V_{\rm CC})\,.
\end{equation}

Fig.~\ref{fig:loop-mixing-limit} presents the reinterpretation of the DUNE sensitivity to the effective long-range coupling (see Fig.~\ref{fig:mass-coupling-plane} and Eq.~\ref{equ:eff_coup}) in terms of active-dark neutrino mixing, assuming that the \(Z\text{-}Z_D\) mixing is generated radiatively via the neutrino loop shown in Fig.~\ref{fig:U1(D)-interaction-loop}. To obtain this reinterpretation, we replace the mass-mixing parameter \(\epsilon^\prime\) with \(\epsilon_{\rm loop}\) and translate the constraints on \(g_{\rm eff}\) into limits on the active-dark mixing using
\begin{equation}
 |U_{\alpha 4}|^2 =
 \left(
 g^2_{\rm eff}
 \times \frac{4 c_W m^2_Z}{g}
 \times \frac{8 \pi^2}{g_D^2 m_4^2}
 \right)^{1/2}\,.
\end{equation}

For illustration, the translation is performed assuming \(g_D = 1\) and \(m_4 = 100\) MeV. We find that DUNE is capable of probing active-sterile neutrino mixing well below existing experimental bounds across a broad range of mediator masses.

\end{widetext}
\bibliographystyle{utcaps_mod}
\bibliography{ref.bib}
\end{document}